\documentclass{llncs}

\usepackage[utf8]{inputenc}
\usepackage[scaled=0.8]{beramono}
\usepackage[T1]{fontenc}
\usepackage{booktabs} 
\usepackage{multirow} 
\usepackage[pagebackref=true]{hyperref} 
\usepackage{makeidx}  
\usepackage{xcolor}   
\usepackage{xspace}   
\usepackage{wrapfig}  
\usepackage{amsmath}
\usepackage{tikz}
\renewcommand*{\backref}[1]{}
\renewcommand*{\backrefalt}[4]{({%
    \ifcase #1 Not cited.%
          \or Cited on p.~#2.%
          \else Cited on pp.~#2.%
    \fi%
    })}

\definecolor{highlight1}{rgb}{0.863,0.863,0.863}
\definecolor{highlight2}{rgb}{0.734,1.000,0.898}
\definecolor{highlight3}{rgb}{0.867,0.758,1.000}
\definecolor{highlight4}{rgb}{1.000,0.977,0.734}
\definecolor{highlight5}{rgb}{1.000,0.816,0.734}

\definecolor{darkblue}{rgb}{0,0,0.4}
\definecolor{lightgray}{rgb}{0.93,0.93,0.93}

\hypersetup{
	colorlinks=true,
	urlcolor=darkblue,
	linkcolor=darkblue,
	citecolor=darkblue,
	pdftitle={Anomalous keys in Tor relays},
	pdfauthor={George Kadianakis, Claudia V. Roberts, Laura M. Roberts, and Philipp Winter},
	pdfkeywords={Tor, RSA, onion service, cryptography, factorization},
	pdfsubject={An analysis of archived RSA public keys of Tor relays},
}

\newcommand{\eg}{{e.g.}\xspace}
\newcommand{\ea}{{et al.}\xspace}
\newcommand{\etc}{{etc.}\xspace}
\newcommand{\first}{(\emph{i})\xspace}
\newcommand{\second}{(\emph{ii})\xspace}
\newcommand{\third}{(\emph{iii})\xspace}
\newcommand{\fourth}{(\emph{iv})\xspace}

\newcommand{\hlfpr}[2]{%
	\texttt{\setlength{\fboxsep}{0pt}%
	\colorbox{highlight1}{\strut #1}#2}}

\begin{document}

\title{``Major key alert!''\\Anomalous keys in Tor relays}
\titlerunning{Anomalous keys in Tor relays}

\author{
	George Kadianakis\inst{1}\thanks{All four authors contributed substantially
	and share first authorship. The names are ordered alphabetically.} \and
	Claudia V. Roberts\inst{2} \and
	Laura M. Roberts\inst{2} \and \\
	Philipp Winter\inst{2}
}

\institute{
	The Tor Project
	\and
	Princeton University
}

\authorrunning{Kadianakis et al.} 

\pagestyle{headings}  

\maketitle

\begin{abstract}
In its more than ten years of existence, the Tor network has seen hundreds of
thousands of relays come and go.  Each relay maintains several RSA keys,
amounting to millions of keys, all archived by The Tor Project.
In this paper, we analyze 3.7 million RSA public keys of Tor relays.  We \first
check if any relays share prime factors or moduli, \second identify relays that
use non-standard exponents, \third characterize malicious relays that we
discovered in the first two steps, and \fourth develop a tool that can determine
what onion services fell prey to said malicious relays.
Our experiments revealed that ten relays shared moduli and 3,557
relays---almost all part of a research project---shared prime factors, allowing
adversaries to reconstruct private keys.  We further discovered 122 relays that
used non-standard RSA exponents, presumably in an attempt to attack onion
services.  By simulating how onion services are positioned in Tor's distributed
hash table, we identified four onion services that were targeted by these
malicious relays.
Our work provides both The Tor Project and onion service operators with tools to
identify misconfigured and malicious Tor relays to stop attacks before they
pose a threat to Tor users.

\end{abstract}

\section{Introduction}
Having seen hundreds of thousands of relays come and go over the last decade,
the Tor network is the largest volunteer-run anonymity network.  To implement
onion routing, all the relays maintain several RSA key pairs, the most important
of which are a medium-term key that rotates occasionally and a long-term key
that ideally never changes.  Most relays run The Tor Project's reference C
implementation on dedicated Linux systems, but some run third-party
implementations or operate on constrained systems such as Raspberry Pis which
raises the question of whether these machines managed to generate safe keys upon
bootstrapping.  Past work has investigated the safety of keys in TLS and SSH
servers~\cite{Heninger2012a}, in nation-wide databases~\cite{Bernstein2013a}, as
well as in POP3S, IMAPS, and SMTPS servers~\cite{Hastings2016a}.  In this work,
we study the Tor network and pay particular attention to Tor-specific aspects
such as onion services.

Relays with weak cryptographic keys can pose a significant threat to Tor users.
The exact impact depends on the type of key that is vulnerable.  In the best
case, an attacker only manages to compromise the TLS layer that protects Tor
cells, which are also encrypted.  In the worst case, an attacker compromises a
relay's long-term ``identity key,'' allowing her to impersonate the relay.  To
protect Tor users, we need methods to find relays with vulnerable keys and
remove them from the network before adversaries can exploit them.

Drawing on a publicly-archived dataset of 3.7 million RSA public
keys~\cite{collector}, we set out to analyze these keys for weaknesses and
anomalies: we looked for shared prime factors, shared moduli, and non-standard
RSA exponents.  To our surprise, we found more than 3,000 keys with shared prime
factors, most belonging to a 2013 research project~\cite{Biryukov2013a}.  Ten
relays in our dataset shared a modulus, suggesting manual interference with the
key generation process.  Finally, we discovered 122 relays whose RSA exponent
differed from Tor's hard-coded exponent.  Most of these relays were meant to
manipulate Tor's distributed hash table (DHT) in an attempt to attack onion
services as we discuss in Section~\ref{sec:targeted-onion-services}.  To learn
more, we implemented a tool---itos\footnote{The name is an acronym for
``identifying targeted onion services.''}---that simulates how onion services
are placed on the DHT, revealing four onion services that were targeted by some
of these malicious relays.  Onion service operators can use our tool to monitor
their services' security; \eg, a newspaper can make sure that its SecureDrop
deployment---which uses onion services---is safe~\cite{securedrop}.

The entities responsible for the incidents we uncovered are as diverse as the
incidents themselves: researchers, developers, and actual adversaries were all
involved in generating key anomalies.  By looking for information that relays
had in common, such as similar nicknames, IP address blocks, uptimes, and port
numbers, we were able to group the relays we discovered into clusters that were
likely operated by the same entities, shedding light on the anatomy of
real-world attacks against Tor.

We publish all our source code and data, allowing third parties such as The Tor
Project to continuously check the keys of new relays and alert developers if
any of these keys are vulnerable or non-standard.\footnote{Our project page is
available online at \url{https://nymity.ch/anomalous-tor-keys/}.}  Tor
developers can then take early action and remove these relays from the network
before adversaries get the chance to take advantage of them.  In summary, we
make the following three contributions:
\begin{itemize}
	\item We analyze a dataset consisting of 3.7 million RSA public keys for
		weak and non-standard keys, revealing thousands of affected keys.

	\item We characterize the relays we discovered, show that many were
		likely operated by a single entity, and uncover four onion services that
		were likely targeted.

	\item Given a set of malicious Tor relays that served as ``hidden service
		directories,'' we develop and implement a method that can reveal what
		onion services these relays were targeting.
\end{itemize}

The rest of this paper details our project.  In Section~\ref{sec:background}, we
provide background information, followed by Section~\ref{sec:related} where we 
discuss related work.  In Section~\ref{sec:method}, we describe our method,
and Section~\ref{sec:results} presents our results.  We discuss our work in
Section~\ref{sec:discussion} and conclude in Section~\ref{sec:conclusion}.

\section{Background}
\label{sec:background}
We now provide brief background on the RSA cryptosystem, how the Tor
network employs RSA, and how onion services are implemented in the Tor network.

\subsection{The RSA cryptosystem}
The RSA public key cryptosystem uses key pairs consisting of a public encryption
key and a privately held decryption key \cite{Rivest1978a}. The encryption key,
or ``RSA public key,'' is comprised of a pair of positive integers: an exponent
$e$ and a modulus $N$. The modulus $N$ is the product of two large, random prime
numbers $p$ and $q$. The corresponding decryption key, or ``RSA private key,''
is comprised of the positive integer pair $d$ and $N$, where $N = pq$ and $d =
e^{-1}$ mod $(p - 1)(q - 1)$.  The decryption exponent $d$ is efficient to
compute if $e$ and the factorization of $N$ are known.

The security of RSA rests upon the difficulty of factorizing $N$ into its prime
factors $p$ and $q$.  While factorizing $N$ is impractical given sufficiently
large prime factors, the greatest common divisor (GCD) of \emph{two moduli} can
be computed in mere microseconds.  Consider two distinct RSA moduli $N_1 = pq_1$
and $N_2 = pq_2$ that share the prime factor $p$.  An attacker could quickly and
easily compute the GCD of $N_1$ and $N_2$, which will be $p$, and then divide the
moduli by $p$ to determine $q_1$ and $q_2$, thus compromising the private key of
both key pairs.  Therefore, it is crucial that both $p$ and $q$ are determined
using a strong random number generator with a unique seed.

Even though the naive GCD algorithm is very efficient, our dataset consists of
more than 3.7 million keys and naively computing the GCD of every pair would
take more than three years of computation (assuming 15 $\mu$s per pair).
Instead, we use the fast pairwise GCD algorithm by Bernstein~\cite{Bernstein04}
which can perform the computation at hand in just a few minutes.

\subsection{The Tor network}
\label{sec:tor-network}
The Tor network is among the most popular tools for digital privacy and
anonymity. As of December 2017, the Tor network consists of around 7,000
volunteer-run relays~\cite{tormetrics}.  Each hour, information about all
relays\footnote{This information includes IP addresses, ports, version numbers,
and cryptographic information, just to name a few.} is summarized in the
\emph{network consensus}, which is used by clients to bootstrap a connection to
the Tor network.  The network consensus is produced by eight
geographically-distributed \emph{directory authorities}, machines run by
individuals trusted by The Tor Project.  For each relay in the consensus, there
is a pointer to its \emph{descriptor}, which contains additional,
relay-specific information such as cryptographic keys.

Each of the $\sim$7,000 relays maintains RSA, Curve25519, and Ed25519 key pairs
to authenticate and protect client traffic~\cite[\S~1.1]{torspec}. In this work,
we analyze the RSA keys.  We leave the analysis of the other key types for
future work.  Each Tor relay has the following three 1024-bit RSA keys:

\begin{figure}[t]
\centering
\tikzset{>=latex}
\begin{tikzpicture}

\node[circle,draw,minimum size=1cm] (R1) at (0, 0) {$R_n$};
\node[circle,draw,minimum size=1cm] (R2) at (5, 0) {$R_{n+1}$};

\draw[<->] (R1.east) -- node [midway, fill=white] (line) {\phantom{TLS layer}} (R2.west);

\node[align=center] at (line) {{\color{gray} Application data}\\1--3 onion
	layers\\TLS layer\\{\color{gray}TCP}\\{\color{gray} IP}};

\end{tikzpicture}
\caption{The protocol stack between two Tor relays $R_n$ and $R_n+1$.  The
	lowest encryption layer is a TLS connection that contains one (between the
	middle and exit relay) to three (between the client and guard relay) onion
	layers.  The onion layers protect the application data that the client is
	sending.} \label{fig:protostack}
\end{figure}
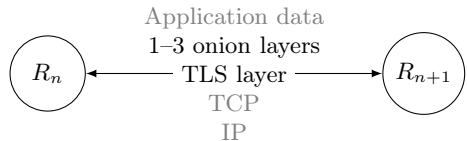

\begin{description}
	\item[Identity key] Relays have a long-term identity key that they use only
		to sign documents and certificates.  Relays are frequently referred to
		by their fingerprints, which is a hash over their identity key.  The
		compromise of an identity key would allow an attacker to impersonate a
		relay by publishing spoofed descriptors signed by the compromised
		identity key.

	\item[Onion key] Relays use medium-term onion keys to decrypt cells when
		circuits are created.  The onion key is only used in the Tor
		Authentication Protocol that is now superseded by the ntor
		handshake~\cite{Goldberg2013a}.  A compromised onion key allows the
		attacker to read the content of cells until the key pair rotates, which
		happens after 28 days~\cite[\S~3.4.1]{dir-spec}.
		However, the onion key layer is protected by a TLS layer (see
		Figure~\ref{fig:protostack}) that an attacker may have to find a way
		around.

	\item[Connection key] The short-term connection keys protect the connection
		between relays using TLS and are rotated at least once a
		day~\cite[\S~1.1]{torspec}.  The TLS connection provides defense in
		depth as shown in Figure~\ref{fig:protostack}.  If compromised, an
		attacker is able to see the encrypted cells that are exchanged between
		Tor relays.
\end{description}

In our work we consider the identity keys and onion keys that each relay has
because the Tor Project has been archiving the public part of the identity and
onion keys for more than ten years, allowing us to draw on a rich
dataset~\cite{collector}. The Tor Project does not archive the connection keys
because they have short-term use and are not found in the network consensus or
relay descriptors.

\subsection{Onion services}
In addition to client anonymity, the Tor network allows operators to set up
anonymous servers, typically called ``onion services.''\footnote{The term ``hidden
services'' was used in the past but was discontinued, in part because onion
services provide more than just ``hiding'' a web site.} The so-called ``hidden
service directories,'' or ``HSDirs,'' are a subset of all Tor relays and
comprise a distributed hash table (DHT) that stores the information necessary
for a client to connect to an onion service.  These HSDirs are a particularly
attractive target to adversaries because they get to learn about onion services
that are set up in the Tor network.  An onion service's position in the DHT is
governed by the following equations:

\begin{equation}
\begin{split}
\textit{secret-id-part} = \textit{SHA-1}(& \textit{time-period} \mid \\
                                         & \textit{descriptor-cookie} \mid \\
                                         & \textit{replica}) \\
\textit{descriptor-id} =  \textit{SHA-1}(& \textit{permanent-id} \mid \\
                                         & \textit{secret-id-part})
\end{split}
\end{equation}

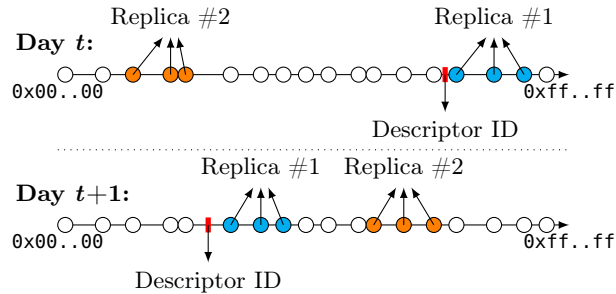
\begin{figure}[t]
\centering
\tikzset{>=latex}
\begin{tikzpicture}

\node at (0.2,0.4) {\textbf{Day \emph{t}+1:}};

\draw[,->] (0,0) node[anchor=north] {{\footnotesize \texttt{0x00..00}}} --
	(6.8,0) node[anchor=north] {{\footnotesize \texttt{0xff..ff}}};

\draw[fill=white] (0.1,0) circle (1mm);
\draw[fill=white] (0.6,0) circle (1mm);
\draw[fill=white] (1.0,0) circle (1mm);
\draw[fill=white] (1.5,0) circle (1mm);
\draw[fill=white] (1.7,0) circle (1mm);

\draw[line width=0.7mm, red] (2,-0.1) -- (2,0.1);
\draw[,->] (2,0) -- (2,-0.5) node[anchor=north] {Descriptor ID};

\draw[fill=cyan] (2.3,0) circle (1mm);
\draw[fill=cyan] (2.7,0) circle (1mm);
\draw[fill=cyan] (3.0,0) circle (1mm);
\draw[,->] (2.3,0) -- (2.6,0.5);
\draw[,->] (2.7,0) -- (2.7,0.5) node[anchor=south] {Replica \#1};
\draw[,->] (3.0,0) -- (2.8,0.5);

\draw[fill=white] (3.3,0) circle (1mm);
\draw[fill=white] (3.6,0) circle (1mm);
\draw[fill=white] (4.0,0) circle (1mm);

\draw[fill=orange] (4.2,0) circle (1mm);
\draw[fill=orange] (4.6,0) circle (1mm);
\draw[fill=orange] (5.0,0) circle (1mm);
\draw[,->] (4.2,0) -- (4.5,0.5);
\draw[,->] (4.6,0) -- (4.6,0.5) node[anchor=south] {Replica \#2};
\draw[,->] (5.0,0) -- (4.7,0.5);

\draw[fill=white] (5.3,0) circle (1mm);
\draw[fill=white] (5.8,0) circle (1mm);
\draw[fill=white] (6.2,0) circle (1mm);
\draw[fill=white] (6.5,0) circle (1mm);

\draw[dotted] (0,1.0) -- (6.8,1.0);

\node at (0.2,2.4) {\textbf{Day \emph{t}:\phantom{+1}}};

\draw[,->] (0,2) node[anchor=north] {{\footnotesize \texttt{0x00..00}}} --
(6.8,2) node[anchor=north] {{\footnotesize \texttt{0xff..ff}}};

\draw[fill=white] (0.1,2) circle (1mm);
\draw[fill=white] (0.6,2) circle (1mm);

\draw[fill=orange] (1.0,2) circle (1mm);
\draw[fill=orange] (1.5,2) circle (1mm);
\draw[fill=orange] (1.7,2) circle (1mm);
\draw[,->] (1.0,2) -- (1.4,2.5);
\draw[,->] (1.5,2) -- (1.5,2.5) node[anchor=south] {Replica \#2};
\draw[,->] (1.7,2) -- (1.6,2.5);

\draw[fill=white] (2.3,2) circle (1mm);
\draw[fill=white] (2.7,2) circle (1mm);
\draw[fill=white] (3.0,2) circle (1mm);
\draw[fill=white] (3.3,2) circle (1mm);
\draw[fill=white] (3.6,2) circle (1mm);
\draw[fill=white] (4.0,2) circle (1mm);
\draw[fill=white] (4.2,2) circle (1mm);
\draw[fill=white] (4.6,2) circle (1mm);
\draw[fill=white] (5.0,2) circle (1mm);

\draw[line width=0.7mm, red] (5.15,1.9) -- (5.15,2.1);
\draw[,->] (5.15,2) -- (5.15,1.5) node[anchor=north] {Descriptor ID};

\draw[fill=cyan] (5.3,2) circle (1mm);
\draw[fill=cyan] (5.8,2) circle (1mm);
\draw[fill=cyan] (6.2,2) circle (1mm);
\draw[,->] (5.3,2) -- (5.7,2.5);
\draw[,->] (5.8,2) -- (5.8,2.5) node[anchor=south] {Replica \#1};
\draw[,->] (6.2,2) -- (5.9,2.5);

\draw[fill=white] (6.5,2) circle (1mm);

\end{tikzpicture}
\caption{Each day, an onion service places its descriptor ID at a pseudorandom
	location in Tor's ``hash ring,'' which consists of all HSDir relays
	(illustrated as circles).}
\label{fig:hash-ring}
\end{figure}

\textit{Secret-id-part} depends on three variables: \textit{time-period}
represents the number of days since the Unix epoch; \textit{descriptor-cookie}
is typically unused and hence empty; and \textit{replica} is set to both the
values 0 and 1, resulting in two hashes for \textit{secret-id-part}.  The
concatenation of both \textit{permanent-id} (the onion service's hashed public
key) and \textit{secret-id-part} is hashed, resulting in \textit{descriptor-id},
which determines the position in the DHT.  When arranging all HSDirs by their
fingerprint in ascending order, the three immediate HSDir neighbors in the positive
direction constitute the first replica while the second replica is at another,
pseudorandom location, as shown in Figure~\ref{fig:hash-ring}.  The onion
service's descriptor ID and hence, its two replicas, changes every day when
\textit{time-period} increments.

\section{Related work}
\label{sec:related}
In 2012, Lenstra \ea~\cite{Lenstra2012a} and Heninger \ea~\cite{Heninger2012a}
independently analyzed a large set of RSA public keys used for TLS, SSH, and
PGP.  Both groups discovered that many keys shared prime factors, allowing an
attacker to efficiently compute the corresponding private keys.  The researchers
showed that the root cause was weak randomness at the time of key generation: 
Many Internet-connected devices lack entropy sources, resulting in predictable
keys.

One year later, Bernstein \ea~\cite{Bernstein2013a} showed similar flaws in
Taiwan's national ``Citizen Digital Certificate'' database.  Among more than two
million 1024-bit RSA keys, the authors discovered 184 vulnerable keys, 103 of
which shared prime factors.  The authors could break the remaining 81 keys by
applying a Coppersmith-type partial-key-recovery
attack~\cite{Coppersmith1996a,Coppersmith1997a}.

Valenta \ea~\cite{Valenta2016a} optimized popular implementations for integer
factorization, allowing them to factor 512-bit RSA public keys on Amazon EC2 in
under four hours for only \$75.  The authors then moved on to survey the RSA key
sizes that are used in popular protocols such as HTTPS, DNSSEC, and SSH,
discovering numerous keys of only 512 bits.

Most recently, in 2016, Hastings \ea~\cite{Hastings2016a} revisited the problem
of weak keys and investigated how many such keys were still on the Internet
four years after the initial studies.  The authors found that many vendors and
device owners never patched their vulnerable devices.  Surprisingly, the number
of vulnerable devices has actually \emph{increased} since 2012.

\section{Method}
\label{sec:method}
In this section, we discuss how we drew on a publicly-available dataset
(Section~\ref{sec:data-collection}) and used Heninger and Halderman's 
fastgcd~\cite{fastgcd} tool to analyze the public keys that 
we extracted from this dataset (Section~\ref{sec:vulnerable-keys}).

\subsection{Data collection}
\label{sec:data-collection}
The Tor Project archives data about Tor relays on its CollecTor
platform~\cite{collector}, allowing researchers to learn what relays were online
at any point in the past.  Drawing on this data source, we compiled a set of RSA
keys by downloading all server descriptors from December 2005 to December 2016 
and extracting the identity and onion keys with the Stem Python
library~\cite{stem}.  Table~\ref{tab:dataset} provides an overview of the
resulting dataset---approximately 200 GB of unzipped data.  Our 3.7 million
public keys span eleven years and were created on one million IP addresses.

Our dataset also contains the keys of Tor's directory authorities.  The
authorities' keys are particularly sensitive: If an attacker were to compromise
more than half of these keys, she could create a malicious network
consensus---which could consist of attacker-controlled relays only---that would
then be used by Tor clients.  Therefore these keys are paramount to the security
of the Tor network.

\begin{table}[t]
	\caption{An overview of our RSA public key dataset.}
	\label{tab:dataset}
	\centering
	\begin{tabular}{l r}
	\toprule

	First key published & 2005-12 \\
	Last key published & 2016-12 \\

	\midrule

	Number of relays (by IP address) & 1,083,805 \\
	Number of onion keys & 3,174,859 \\
	Number of identity keys & 588,945 \\
	Total number of public keys & 3,763,804 \\

	\bottomrule
	\end{tabular}
\end{table}

\subsection{Finding vulnerable keys}
\label{sec:vulnerable-keys}
To detect weak, potentially factorable keys in the Tor network, we used Heninger
and Halderman's fastgcd~\cite{fastgcd} tool which takes as input a set
of moduli from public keys and then computes the pair-wise greatest common
divisor of these moduli.  Fastgcd's C implementation is based on a
quasilinear-time algorithm for factoring a set of integers into their co-primes.
We used the PyCrypto library~\cite{pycrypto} to turn Tor's PKCS\#1-padded,
PEM-encoded keys into fastgcd's expected format, which is hex-encoded
moduli.  Running fastgcd over our dataset took less than 20 minutes on
a machine with dual, eight-core 2.8 GHz Intel Xeon E5 2680 v2 processors with
256 GB of RAM.

Fastgcd benefits from having access to a pool of moduli that is as large as possible
because it allows the algorithm to draw on a larger factor base to use on each
key~\cite{Heninger2012a}.  To that end, we reached out to Heninger's group at
the University of Pennsylvania, and they graciously augmented their 129 million
key dataset with our 3.6 million keys and subsequently searched for shared
factors.  The number of weak keys did not go up, but this experiment gave us
more confidence that we had not missed weak keys.

\section{Results}
\label{sec:results}
We present our results in four parts, starting with shared prime factors
(Section~\ref{sec:shared-primes}), followed by shared moduli
(Section~\ref{sec:shared-moduli}), unusual exponents 
(Section~\ref{sec:unusual-exponents}), and finally, targeted onion services
(Section~\ref{sec:targeted-onion-services}).

\subsection{Shared prime factors}
\label{sec:shared-primes}
Among all 588,945 identity keys, fastgcd found that 3,557 (0.6\%)
moduli share prime factors.  We believe that 3,555 of these keys were all
controlled by a single research group, and upon contacting the authors of the
Security \& Privacy 2013 paper entitled ``Trawling for Tor hidden
services''~\cite{Biryukov2013a}, we received confirmation that these relays
were indeed run by their research group.  The authors informed us that the weak
keys were caused by a shortcoming in their key generation tool. The issue
stemmed from the fact that their tool first generated thousands of prime numbers
and then computed multiple moduli using combinations of those prime numbers in a
greedy fashion without ensuring that the same primes were not reused.  Because
of the following shared properties, we are confident that all relays were
operated by the researchers:

\begin{enumerate}
	\item All relays were online either between November 11, 2012 and
		November 16, 2012 or between January 14, 2013 and February 6, 2013,
		suggesting two separate experiments. We verified this by checking how
		long the relays stayed in the Tor network consensus. The Tor consensus
		is updated hourly and documents which relays are available at a
		particular time. This data is archived by The Tor Project and is made
		publicly available on the CollecTor platform~\cite{collector}.

	\item All relays exhibited a predictable port assignment scheme.  In
		particular, we observed ports \{7003, 7007, \dots, 7043, 7047\} and
		\{8003, 8007, \dots, 8043, 8047\}.

	\item Except for two machines that were located in Russia and Luxembourg,
		all machines were hosted in Amazon's EC2 address space.  All machines
		except the one located in Luxembourg used Tor version 0.2.2.37.

	\item All physical machines had multiple fingerprints.  1,321 of these 3,557
		relays were previously characterized by Winter
		\ea~\cite[\S~5.1]{Winter2016a}.
\end{enumerate}

The remaining two keys belonged to a relay named ``Desaster\-Blaster,'' whose
origins we could not determine. Its router descriptor indicates that the relay
has been hosted on a MIPS machine which might suggest an embedded device with a
weak random number generator:

{\footnotesize
\begin{verbatim}
router DesasterBlaster 62.226.55.122 9001 0 0
platform Tor 0.2.2.13-alpha on Linux mips
\end{verbatim}
}

To further investigate, we checked whether the relay ``Desaster\-Blaster'' shares
prime factors with any other relays. It appears that the relay has rotated
multiple identity keys, and it only shares prime factors with its own keys.
Unfortunately the relay did not have any contact information configured which is
why we could not get in touch with its operator.

\subsection{Shared moduli}
\label{sec:shared-moduli}
In addition to finding shared prime factors, we discovered relays that share a
\emph{modulus}, giving them the ability to calculate each other's private keys.
With $p$, $q$, and each other's $e$s in hand, the two parties can compute
each other's decryption exponent $d$, at which point both parties now know the
private decryption keys.

Table~\ref{tab:moduli} shows these ten relays with shared moduli clustered into
four groups. The table shows the relays' truncated, four-byte fingerprint, IP
addresses, and RSA exponents.  Note that the Tor client hard-codes the RSA
exponent to 65,537~\cite[\S~0.3]{torspec}, a recommended value that is resistant
to attacks against low public exponents~\cite[\S~4]{Boneh1999a}.  Any value
other than 65,537 indicates non-standard key generation.  All IP addresses were
hosted by OVH, a popular French hosting provider, and some of the IP addresses
hosted two relays, as our color coding indicates.  Finally, each group shared a
four- or five-digit prefix in their fingerprints.  We believe that a single
attacker controlled all these relays with the intention to manipulate the
distributed hash table that powers onion services~\cite{Biryukov2013a}---the
shared fingerprint prefix is an indication.  Because the modulus is identical,
we suspect that the attackers iterated over the relays' RSA exponents to come up
with the shared prefix.  The Tor Project informed us that it discovered and
blocked these relays in August 2014 when they first came online.

\begin{table}[t]
	\caption{Four groups of relays that have a shared modulus.  All relays
	further share a fingerprint prefix in groups of two or three, presumably to
	manipulate Tor's distributed hash table.}
	\label{tab:moduli}

	\centering
	\begin{tabular}{l l r}
	\toprule

	Short fingerprint & IP address & Exponent \\
	\midrule

	\texttt{\setlength{\fboxsep}{0pt}%
	\colorbox{highlight1}{\strut 838A}296A} & 188.165.164.163 &
	1,854,629 \\

	\texttt{\setlength{\fboxsep}{0pt}%
	\colorbox{highlight1}{\strut 838A}305F} &
	{\setlength{\fboxsep}{0pt}\colorbox{highlight3}{\strut 188.165.26.13}} &
	718,645 \\

	\texttt{\setlength{\fboxsep}{0pt}%
	\colorbox{highlight1}{\strut 838A}71E2} &
	{\setlength{\fboxsep}{0pt}\colorbox{highlight2}{\strut 178.32.143.175}} &
	220,955 \\

	\midrule

	\texttt{\setlength{\fboxsep}{0pt}%
	\colorbox{highlight1}{\strut 2249E}B42} &
	{\setlength{\fboxsep}{0pt}\colorbox{highlight3}{\strut 188.165.26.13}} &
	4,510,659 \\

	\texttt{\setlength{\fboxsep}{0pt}%
	\colorbox{highlight1}{\strut 2249E}C78} &
	{\setlength{\fboxsep}{0pt}\colorbox{highlight2}{\strut 178.32.143.175}} &
	1,074,365 \\

	\midrule

	\texttt{\setlength{\fboxsep}{0pt}%
	\colorbox{highlight1}{\strut E1EF}A388} & 188.165.3.63 &
	18,177 \\

	\texttt{\setlength{\fboxsep}{0pt}%
	\colorbox{highlight1}{\strut E1EF}8985} &
	{\setlength{\fboxsep}{0pt}\colorbox{highlight4}{\strut 188.165.138.181}} &
	546,019 \\

	\texttt{\setlength{\fboxsep}{0pt}%
	\colorbox{highlight1}{\strut E1EF}9EB8} &
	{\setlength{\fboxsep}{0pt}\colorbox{highlight5}{\strut 5.39.122.66}} &
	73,389 \\

	\midrule

	\texttt{\setlength{\fboxsep}{0pt}%
	\colorbox{highlight1}{\strut 410B}A17E} &
	{\setlength{\fboxsep}{0pt}\colorbox{highlight4}{\strut 188.165.138.181}} &
	1,979,465 \\

	\texttt{\setlength{\fboxsep}{0pt}%
	\colorbox{highlight1}{\strut 410B}B962} &
	{\setlength{\fboxsep}{0pt}\colorbox{highlight5}{\strut 5.39.122.66}} &
	341,785 \\

	\bottomrule
	\end{tabular}
\end{table}

\subsection{Unusual exponents}
\label{sec:unusual-exponents}
Having accidentally found a number of relays with non-standard exponents in
Section~\ref{sec:shared-moduli}, we checked if our dataset featured more relays
with exponents other than 65,537.  Non-standard exponents may indicate that a
relay was after a specific fingerprint in order to position itself in Tor's hash
ring.  To obtain a fingerprint with a given prefix, an adversary repeatedly has
to modify any of the underlying key material $p$, $q$, or $e$ until they result
in the desired prefix.  Repeated modification of $e$ is significantly more
efficient than modifying $p$ or $q$ because it is costly to verify if a large
number is prime.  Leveraging this method, the tool Scallion~\cite{scallion}
generates vanity onion service domains by iterating over the service's public
exponent.

Among all of our 3.7 million keys, 122 possessed an exponent other than 65,537.
One relay had both non-standard identity \emph{and} onion key exponents while
all remaining relays only had non-standard identity key exponents.  Ten of these
relays further had a shared modulus, which we discuss in
Section~\ref{sec:shared-moduli}.  Assuming that these relays positioned
themselves in the hash ring to attack an onion service, we wanted to find out
what onion services they targeted.  One can identify the victims by first
compiling a comprehensive list of onion services and then determining each
service's position in the hash ring at the time the malicious HSDirs were
online.

\subsection{Identifying targeted onion services}
\label{sec:targeted-onion-services}

We obtained a list of onion services by augmenting the list of the Ahmia
search engine~\cite{ahmia} with services that we discovered via Google searches
and by contacting researchers who have done similar work~\cite{Matic2015a}.  We
ended up with a list of 17,198 onion services that were online at some point in
time.  Next, we developed a tool that takes as input our list of onion services
and the malicious HSDirs we discovered.\footnote{Both the tool and our list of
onion services are available online at
\url{https://nymity.ch/anomalous-tor-keys/}.} The tool then calculates all
descriptors these onion services ever generated and checks if any HSDir shared
five or more hex digits in its fingerprint prefix with the onion service's
descriptor.  We chose the threshold of five manually because it is unlikely to
happen by chance yet easy to create a five-digit collision.

It is difficult to identify all targeted onion services because \first our list
of onion services does not tell us when a service was online, \second an HSDir
could be responsible for an onion service simply by chance rather than on
purpose, resulting in a false positive, and \third our list of onion services is
not exhaustive, so we are bound to miss victims.  Nevertheless our tool
identified four onion services (see Table~\ref{tab:targeted}) for which we have
strong evidence that they were purposely targeted.  While HSDirs are frequently
in the vicinity of an onion service's descriptor by accident, the probability of
being in its vicinity for several days in a row or cover both replicas by chance
is negligible.  Table~\ref{tab:collisions} shows all partial collisions in
detail.  Because none of these four services seem to have been intended for
private use, we are comfortable publishing them.

\begin{table}[t]
	\caption{The four onion services that were most likely targeted at some
	point.  The second column indicates if only one or both replicas were
	attacked while the third column shows the duration of the attack.}
	\label{tab:targeted}
	\centering
	\begin{tabular}{l c l}
	\toprule
	Onion service & Replicas & Attack duration \\
	\midrule
	\texttt{22u75kqyl666joi2.onion} & 2 & Two consecutive days \\
	\texttt{n3q7l52nfpm77vnf.onion} & 2 & Six non-consecutive days \\
	\texttt{silkroadvb5piz3r.onion} & 1 & Nine mostly consecutive days \\
	\texttt{thehub7gqe43miyc.onion} & 2 & One day \\
	\bottomrule
	\end{tabular}
\end{table}

\begin{description}
	\item[\texttt{22u75kqyl666joi2.onion}] The service appears to be offline
		today, so we were unable to see for ourselves what it hosted.  According
		to cached index pages we found online, the onion service used to host a
		technology-focused forum in Chinese.  A set of relays targeted the onion
		service on both August 14 and 15, 2015 by providing nine out of the
		total of twelve responsible HSDirs.

	\item[\texttt{n3q7l52nfpm77vnf.onion}] As of February 2017, the service is
		still online, hosting the ``Marxists Internet Archive,'' an online
		archive of literature.\footnote{The onion service seems to be identical
		to the website \url{https://www.marxists.org} (visited on 2017-05-09).}
		A set of relays targeted the onion service from November 27 to December
		4, 2016.  The malicious HSDirs acted inconsistently, occasionally
		targeting only one replica.

	\item[\texttt{silkroadvb5piz3r.onion}] The onion service used to host the
		Silk Road marketplace, whose predominant use was a market for narcotics.
		The service was targeted by a set of relays from May 21 to June 3, 2013.
		The HSDirs were part of a measurement experiment that resulted in a blog
		post~\cite{OCearbhaill2013a}.

	\item[\texttt{thehub7gqe43miyc.onion}] The onion service used to host a
		discussion forum, ``The Hub,'' focused on darknet markets.  A set of
		relays targeted both of The Hub's replicas from August 22, 2015.
\end{description}

Our data cannot provide insight into what the HSDirs did once they controlled
the replicas of the onion services they targeted.  The HSDirs could have counted
the number of client requests, refused to serve the onion service's descriptor
to take it offline, or correlate client requests with guard relay traffic in
order to deanonymize onion service visitors as it was done by the CMU/SEI
researchers in 2014~\cite{Dingledine2014a}.  Since these attacks were
short-lived we find it unlikely that they were meant to take offline the
respective onion services.

\section{Discussion}
\label{sec:discussion}
We now turn to the technical and ethical implications of our research, propose
possible future work, and explain how the next generation of onion
services will thwart DHT manipulation attacks.

\subsection{Implications of anomalous Tor keys}
\paragraph{Implications for the network}
As touched on earlier in Section~\ref{sec:tor-network}, the main use of the
identity key in Tor is to sign the relay's descriptor, which includes various
information about the relay, \eg, its IP address and contact information.
Relays publish their public identity keys in their descriptor.  The network
consensus acts as the public key infrastructure of Tor.  Signed by the directory
authorities whose public keys are hard-coded in Tor's source code, the network
consensus points to the descriptors of each Tor relay that is currently online.
If an attacker were to break the identity key of a relay (as we demonstrated),
she could start signing descriptors in the relay's name and publishing them. The
adversary could publish whatever information she wanted in the descriptor, \eg
her own IP address, keys, \etc, in order to fool Tor clients.  In other words,
weak keys allow adversaries to obtain the affected relay's reputation which
matters because Tor clients make routing decisions based on this reputation.

The Tor protocol's use of forward secrecy mitigates the potential harm of weak
keys.  Recall that a relay's long-lived identity keys are only used to sign
data, so forward secrecy does not apply here.  Onion keys, however, are used for
decryption and encryption and are rotated by default every 28
days~\cite[\S~3.4.1]{dir-spec}.  An attacker who manages to compromise a weak
onion key is still faced with the underlying TLS layer, shown in
Figure~\ref{fig:protostack}, which provides defense in depth.  The Tor
specification requires the keys for the TLS layer to be rotated at least once a
day~\cite[\S~1.1]{torspec}, making it difficult to get any use out of
compromised onion keys.

\paragraph{Implications for Tor users}
To understand how Tor users are affected by weak keys we need to distinguish
between \emph{targeting} and \emph{hoovering} adversaries.\footnote{We here use
Jaggard and Syverson's nomenclature of an adversary that either targets
specific Tor users (targeting) or hoovers up all available data to
deanonymize as many users as possible (hoovering)~\cite{Jaggard2017a}.}
The goal of targeting adversaries is to focus an attack on a small number of
users among the large set of all Tor users.  Generally speaking, weak keys can
be problematic in a targeted setting if they allow an attacker to gain access to
a Tor relay she would otherwise be unable to control.  This can be the case if
the attacker learned the targeted user's guard relay, and the guard happens
to have weak keys.  However, judging by our experimental results, the
probability of an attacker's knowing a targeted user's guard relay \emph{and} 
said guard relay's having vulnerable keys is very low.

Hoovering adversaries are opportunistic by definition and seek to deanonymize as
many Tor users as possible.  Recall that Tor clients use a long-lived guard
relay as their first hop and two randomly chosen relays for the next two
hops.\footnote{We refer to these relays as randomly chosen for simplicity, but
the path selection algorithm is more complicated.}  A single compromised relay
is not necessarily harmful to users but weak keys can be a problem if a user
happens to have a guard relay with weak keys \emph{and} selects an exit relay
that also has weak keys, allowing the hoovering adversary to deanonymize the
circuit.  Again, considering the low prevalence of weak keys and the ease with
which The Tor Project could identify and block relays with weak keys, hoovering
adversaries pose no significant threat.

\subsection{Preventing non-standard exponents}
Recall that the Tor reference implementation hard-codes its public RSA exponent
to 65,537~\cite[\S~0.3]{torspec}.  The Tor Project could prevent non-standard
exponents by having the directory authorities reject relays whose descriptors
have an RSA exponent other than 65,537, thus slowing down the search for
fingerprint prefixes.  Adversaries would then have to iterate over the primes
$p$ or $q$ instead of the exponent, rendering the search computationally
more expensive because of the cost of primality tests.  Given that we discovered
only 122 unusual exponents in over ten years of data, we believe that rejecting
non-standard exponents is a viable defense in depth.

\subsection{Analyzing onion service public keys}
Future work should shed light on the public keys of onion services.  Onion
services have an incentive to modify their fingerprints to make them both
recognizable and easier to remember.  Facebook, for example, was lucky to
obtain the easy-to-remember onion domain
\url{facebookcorewwwi.onion}~\cite{facebook}.  The tool Scallion assists onion
service operators in creating such vanity domains~\cite{scallion}.  The
implications of vanity domains on usability and security are still poorly
understood~\cite{vanity-domains}.  Unlike the public keys of relays, onion
service keys are not archived, so a study would have to begin with actively
fetching onion service keys.

\subsection{Investigating the vulnerability of Tor relays to attacks on Diffie-Hellman}
Recent work has demonstrated how Diffie-Hellman key exchange is vulnerable to
attack~\cite{Adrian2015a,Valenta2017a,Dorey2017a}.  Because Tor uses
Diffie-Hellman, we decided to investigate how it might be affected by those
findings.  The Tor specification states that the implementation uses the Second
Oakley Group for Diffie-Hellman, where the prime number is 1024 bits
long~\cite[\S~0.3]{torspec}. To gather evidence of this usage, we performed an
nmap scan on Tor relays using the ssl-dh-params script~\cite{nmapdhscript},
which confirmed the Tor specification. The use of a 1024-bit prime is
concerning because Adrian \ea~\cite{Adrian2015a} stated that ``1024-bit discrete
log may be within reach for state-level actors,'' and thus, they suggest a move
to 2048 bits. The authors also mention that developers should move away from
using 1024-bit RSA, as well, which Tor uses.

\subsection{\textit{In vivo} Tor research}
Caution must be taken when conducting research using the live Tor network.
Section~\ref{sec:shared-primes} showed how a small mistake in key generation led
to many vulnerable Tor relays.  To keep its users safe, The Tor Project has
recently launched a research safety board whose aim is to assist researchers in
safely conducting Tor measurement studies~\cite{safety-board}.  This may entail
running experiments in private Tor networks that are controlled by the
researchers or using network simulators such as Shadow~\cite{Jansen2012a}.

As for our own work, we were in close contact with Tor developers throughout our
research effort and shared preliminary results as we progressed.  Once we wrote
up our findings in a technical report, we brought it to the broader Tor
community's attention by sending an email to the tor-dev mailing
list~\cite{Roberts2017a}.  On top of that, we adopted open science practices and
wrote both our code and paper in the open, allowing interested parties to follow
our progress easily.

\subsection{The effect of next-generation onion services}
As of December 2017, The Tor Project is testing the implementation of
next-generation onion services~\cite{prop224}.  In addition to stronger
cryptographic primitives, the design fixes the issue of predicting an onion
service's location in the hash ring by incorporating a random element.  This
element is produced by having the directory authorities agree on a random number
once a day~\cite{prop250}.  The random number is embedded in the consensus
document and used by clients to fetch an onion service's descriptor.  Attackers
will no longer be able to attack onion services by positioning HSDirs in the
hash ring; while they have several hours to compute a key pair that positions
their HSDirs next to the onion service's descriptor (which is entirely
feasible), it takes at least 96 hours to get the HSDir flag from the directory
authorities~\cite[\S~3.4.2]{dir-spec}, so attackers cannot get the flag in time.
We expect this design change to disincentivize attackers from manipulating their
keys to attack onion services.

\section{Conclusion}
\label{sec:conclusion}

Inspired by recent research that studied weak keys in deployed systems, we set
out to investigate if the Tor network suffers from similar issues.  To that
end, we drew on a public archive containing cryptographic keys of Tor relays
dating back to 2005, which we subsequently analyzed for weak RSA keys.  We
discovered that \first ten relays shared an RSA modulus, \second 3,557 relays
shared prime factors, and \third 122 relays used non-standard RSA exponents.

Having uncovered these anomalies, we then proceeded to characterize the
affected relays, tracing back the issues to mostly harmless experiments run by
academic researchers and hobbyists, but also to attackers that targeted Tor's
distributed hash table which powers onion services.  To learn more, we
implemented a tool that can determine what onion services were attacked by a
given set of malicious Tor relays, revealing four onion services that fell prey
to these attacks.

The practical value of our work is twofold.  First, our uncovering and
characterizing of Tor relays with anomalous keys provides an anatomy of
real-world attacks that The Tor Project can draw upon to improve its monitoring
infrastructure for malicious Tor relays.  Second, our work provides The Tor
Project with tools to verify the RSA keys of freshly set up relays, making the
network safer for its users.  In addition, onion service operators can use our
code to monitor their services and get notified if Tor relays make an effort to
deanonymize their onion service.  We believe that this is particularly useful
for sensitive deployments such as SecureDrop instances.

\section*{Acknowledgements}
We want to thank Nadia Heninger and Josh Fried for augmenting their database
with our moduli and attempting to find factors in them. We also want to
thank Ralf-Philipp Weinmann, Ivan Pustogarov, Alex Biryukov from the Trawling
research team and Donncha O'Cearbhaill from The Tor Project for providing us
with additional information that helped us in our analysis of the weak keys.
Finally, we want to thank Edward W. Felten for providing valuable feedback on
an earlier version of our paper.  This research was supported by the Center for
Information Technology Policy at Princeton University and the National Science
Foundation Awards CNS-1540066, CNS-1602399, CNS-1111539, CNS-1314637,
CNS-1520552, and CNS-1640548.

\bibliographystyle{splncs03}
\bibliography{references.bib}{}

\newpage

\appendix

\section{Potentially targeted onion services}

\begin{table*}

	\caption{The details of the attacks on four onion services.  The second
		column shows the fingerprints of the HSDirs that were participating in
		the attack.  The third column shows the affected onion service
		descriptors, followed by the date of the attack in the last column.}
	\label{tab:collisions}

	\centering
	\scriptsize
	\begin{tabular}{l l l l}
	\toprule
	Onion service & Truncated HSDir fingerprint & Truncated onion service descriptor & Date \\
	\midrule
	\texttt{22u75kqyl666joi2} & \hlfpr{325CAC0}{B7FA8CD77E39D} & \hlfpr{325CAC0}{8B0A3180B590E} & 2015-08-14 \\
	                          & \hlfpr{325CAC0}{AB1AAD27493B9} & \hlfpr{325CAC0}{8B0A3180B590E} & 2015-08-14 \\
	                          & \hlfpr{325CAC0}{A43B2121B81CD} & \hlfpr{325CAC0}{8B0A3180B590E} & 2015-08-14 \\
	                          & \hlfpr{FA25674}{1ED22FD96AF5A} & \hlfpr{FA25674}{0740356704AB8} & 2015-08-14 \\
	                          & \hlfpr{FA25674}{3ACFCA9B7C85D} & \hlfpr{FA25674}{0740356704AB8} & 2015-08-14 \\
	                          & \hlfpr{E5E77832}{6AF0FF0A634A} & \hlfpr{E5E77832}{45096FB554A1} & 2015-08-15 \\
	                          & \hlfpr{A5C59B3}{D0FFBDE88405E} & \hlfpr{A5C59B3}{CD34802FC4AC3} & 2015-08-15 \\
	                          & \hlfpr{A5C59B3}{FCCD2FA8FAD42} & \hlfpr{A5C59B3}{CD34802FC4AC3} & 2015-08-15 \\
	                          & \hlfpr{A5C59B3}{FD5625A0D85D1} & \hlfpr{A5C59B3}{CD34802FC4AC3} & 2015-08-15 \\
	\midrule
	\texttt{n3q7l52nfpm77vnf} & \hlfpr{A0E83AA1}{91220B240EC0} & \hlfpr{A0E83AA1}{15098CA7FE9B} & 2016-11-27 \\
	                          & \hlfpr{A0E83AA}{28382135DC839} & \hlfpr{A0E83AA}{115098CA7FE9B} & 2016-11-27 \\
	                          & \hlfpr{EBF154D}{A21B49101ED5B} & \hlfpr{EBF154D}{809425D3E923E} & 2016-11-28 \\
	                          & \hlfpr{EBF154D8}{BB6EECCC2921} & \hlfpr{EBF154D8}{09425D3E923E} & 2016-11-28 \\
	                          & \hlfpr{EBF154D}{9E2B10A2420E0} & \hlfpr{EBF154D}{809425D3E923E} & 2016-11-28 \\
	                          & \hlfpr{6761D2B}{E758FA0D76822} & \hlfpr{6761D2B}{CF40FF34274F3} & 2016-11-29 \\
	                          & \hlfpr{59E415D}{78921BFF88168} & \hlfpr{59E415D}{5075157CAADB7} & 2016-11-29 \\
	                          & \hlfpr{26597E6}{2875C498AC139} & \hlfpr{26597E6}{048BF7CC9D593} & 2016-11-30 \\
	                          & \hlfpr{26597E6}{1DDFEE78F336D} & \hlfpr{26597E6}{048BF7CC9D593} & 2016-11-30 \\
	                          & \hlfpr{7CDB224}{FE64F2A50CC50} & \hlfpr{7CDB224}{DC51432C037C5} & 2016-11-30 \\
	                          & \hlfpr{2D148D3}{EBF9D2B9D8CCB} & \hlfpr{2D148D3}{CB6C5FC4DCA14} & 2016-12-01 \\
	                          & \hlfpr{2E25D84}{69331FEAE933D} & \hlfpr{2E25D84}{2BF5DDA936BA2} & 2016-12-05 \\
	                          & \hlfpr{2E25D84}{7E579AED1B0EC} & \hlfpr{2E25D84}{2BF5DDA936BA2} & 2016-12-05 \\
	                          & \hlfpr{2E25D84}{54C96E20CF153} & \hlfpr{2E25D84}{2BF5DDA936BA2} & 2016-12-05 \\
	                          & \hlfpr{2E25D84}{6564DCBE43CD2} & \hlfpr{2E25D84}{2BF5DDA936BA2} & 2016-12-05 \\
	                          & \hlfpr{2E25D84}{47518DA93B4FF} & \hlfpr{2E25D84}{2BF5DDA936BA2} & 2016-12-05 \\
	                          & \hlfpr{264EA12}{B47CBCC8043C5} & \hlfpr{264EA12}{410F7D9CD6E54} & 2016-12-05 \\
	                          & \hlfpr{264EA12}{84855A596D5D6} & \hlfpr{264EA12}{410F7D9CD6E54} & 2016-12-05 \\
	                          & \hlfpr{264EA12}{B4C46672E002C} & \hlfpr{264EA12}{410F7D9CD6E54} & 2016-12-05 \\
	\midrule
	\texttt{silkroadvb5piz3r} & \hlfpr{BC89A}{92F53581C4F6169} & \hlfpr{BC89A}{889D3DF7F0027A5} & 2013-05-21 \\
	                          & \hlfpr{712CA}{45AF4055E7AC69A} & \hlfpr{712CA}{3DEF4EB21C76A95} & 2013-05-22 \\
	                          & \hlfpr{DE1529}{9D7EE5882F0BEF} & \hlfpr{DE1529}{316F5172B35B8E} & 2013-05-23 \\
	                          & \hlfpr{FF0BF}{54FBEEE7A003CE6} & \hlfpr{FF0BF}{49076AA63C97FA2} & 2013-05-24 \\
	                          & \hlfpr{E9F25}{C4899F9DC81E48E} & \hlfpr{E9F25}{BBA0D4501FAE18B} & 2013-05-28 \\
	                          & \hlfpr{B81B43}{C015DE42D05208} & \hlfpr{B81B43}{637F22592ECC80} & 2013-05-29 \\
	                          & \hlfpr{59529}{817C6E797D78311} & \hlfpr{59529}{79BD9FEECE847E7} & 2013-05-31 \\
	                          & \hlfpr{BCB332}{864640653892D4} & \hlfpr{BCB332}{36E0AD461DF585} & 2013-06-02 \\
	                          & \hlfpr{51FC17}{8DFF3D0B869760} & \hlfpr{51FC17}{2F0062B623A39D} & 2013-06-03 \\
	\midrule
	\texttt{thehub7gqe43miyc} & \hlfpr{F6961286}{D361F825A9AD} & \hlfpr{F6961286}{C2FEEA8DEDEB} & 2015-08-22 \\
	                          & \hlfpr{F6961286C}{453F6A6381D} & \hlfpr{F6961286C}{2FEEA8DEDEB} & 2015-08-22 \\
	                          & \hlfpr{F6961286}{D826D7D1C0F9} & \hlfpr{F6961286}{C2FEEA8DEDEB} & 2015-08-22 \\
	                          & \hlfpr{816FEE1}{6200BE1719D00} & \hlfpr{816FEE1}{5D26F41A72039} & 2015-08-22 \\
	\bottomrule
	\end{tabular}
\end{table*}

\end{document}